\documentclass[]{revtex4-1}
\usepackage{amsmath}
\usepackage{color}
\usepackage{graphicx}
\usepackage{amssymb}
\usepackage{subfigure}

\begin{document}

\title{Unimpeded tunneling in graphene nanoribbons}

\author{O. Roslyak$^1$, A. Iurov$^1$, Godfrey  Gumbs$^{1,3}$ and Danhong Huang$^2$}
\affiliation{$^1$Department of Physics and Astronomy,
Hunter College of  City University of New York,
695 Park Avenue, New York, NY 10065-50085}
\affiliation{$^2$Air Force Research Laboratory (AFRL/RVSS), Kirtland Air
Force Base, NM 87117, USA}
\affiliation{$^3$Donostia International Physics Center (DIPC),
P. de Manuel Lardizabal, 4, 20018 San Sebasti\'an,
Basque Country, Spain }

\pacs{73.20.At, 73.21.Hb, 73.50.Bk, 73.61.Wp}

\begin{abstract}
We studied the Klein paradox in zigzag (ZNR) and anti-zigzag (AZNR) graphene nanoribbons. Due to the fact that ZNR (the number of lattice sites across the nanoribbon = N is even) and AZNR (N is odd) configurations are indistinguishable when treated by the Dirac equation, we supplemented the model with a pseudo-parity operator whose eigenvalues correctly depend on the sublattice wavefunctions for the number of carbon atoms across the ribbon, in agreement with the tight-binding model. We have shown that the Klein tunneling in zigzag nanoribbons is related to conservation of the pseudo-parity rather than pseudo-spin in infinite graphene. The perfect transmission in the case of head-on incidence is replaced by perfect transmission at the center of the ribbon and the chirality is interpreted as the projection of the pseudo-parity on momentum at different corners of the Brillouin zone.
\end{abstract}
\maketitle

\par
\begin{figure}[htbp]
  \centering
  \subfigure[]{
  \includegraphics[width=0.45\textwidth]{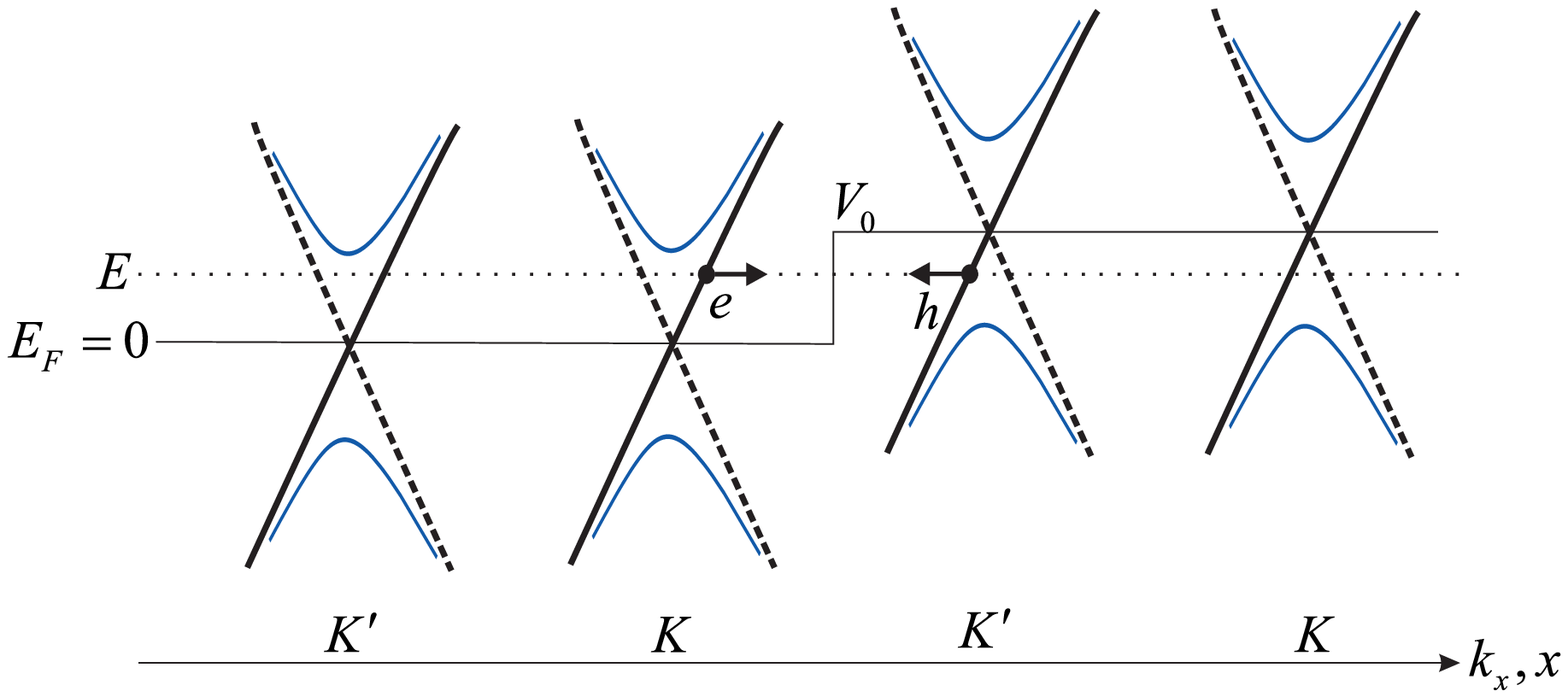}
  }
  \subfigure[]{
  \includegraphics[width=0.45\textwidth]{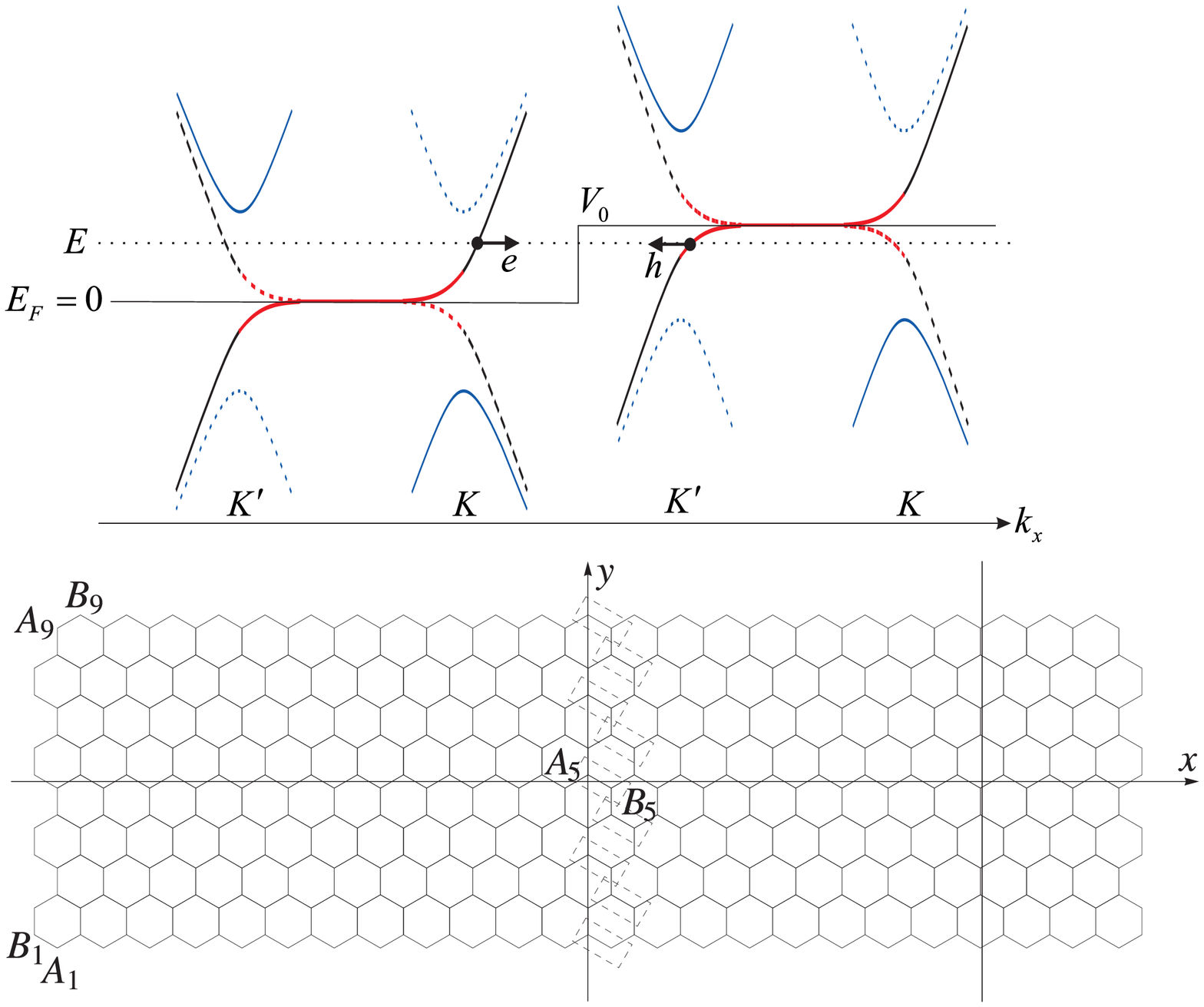}
  }
  \subfigure[]{
  \includegraphics[width=0.45\textwidth]{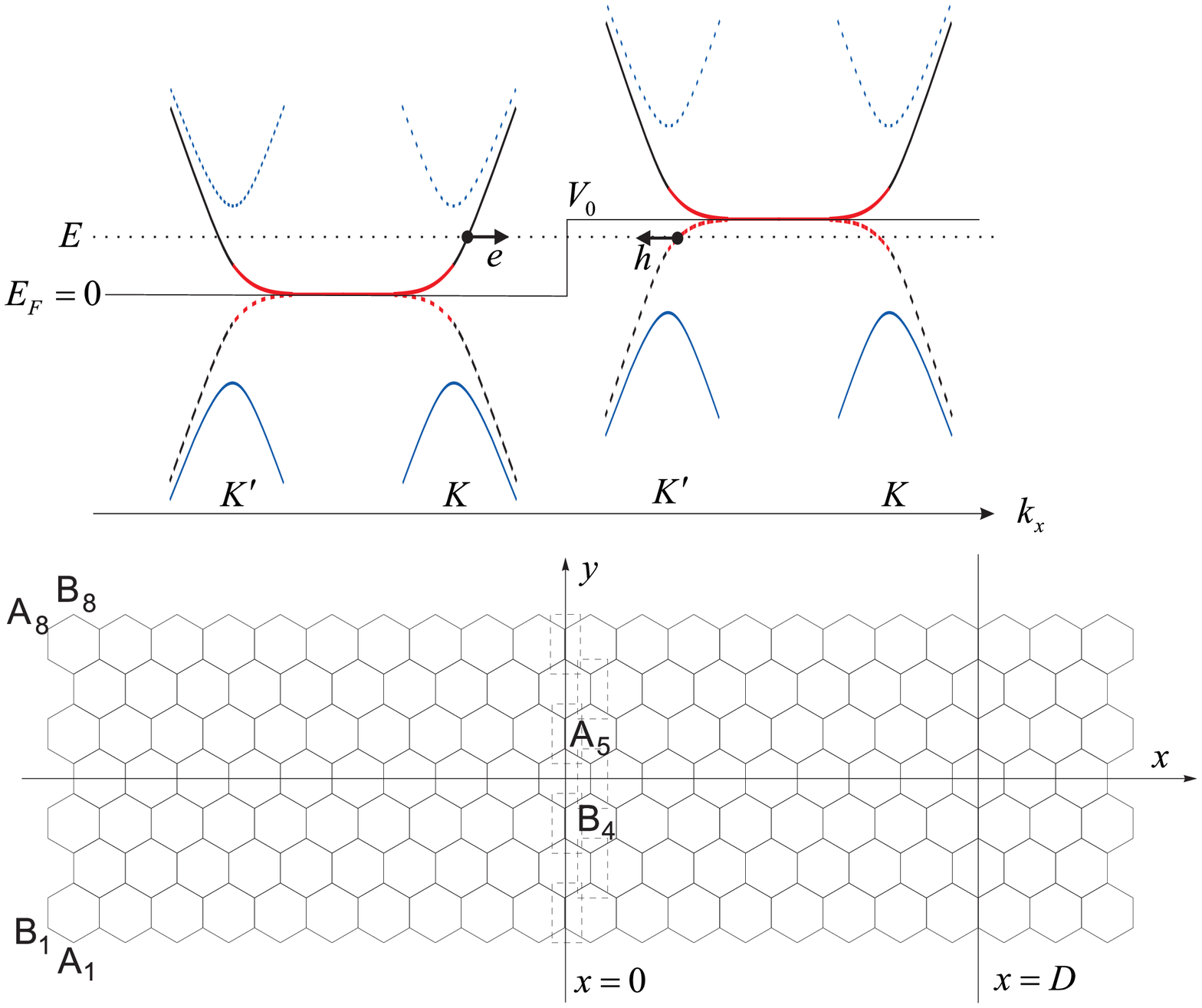}
  }
  \caption{(Color online). Panel (a) shows infinite graphene $n-p-n$ junction and corresponding chiral dispersion curves. The $\pi,\pi^\star$ bands are depicted by black lines; the nearby parabolic shaped (blue) curves denote $\mathcal{V},\mathcal{V}^\star$ bands. The different values of the pseudo-spin projected on the momentum (eigenvalues of $\sim \sigma \cdot p$ given by $s_B p_x /\vert p \vert$) are shown in solid and dashed curves. Panel (b) describes 9-AZNR $n-p-n$ junction and chiral particle dispersion. The two different values of pseudo-parity projection on the pseudo-spin $1 - s^\star \cdot \eta$ are presented in solid (dashed) curves. Panel (c) shows 8-ZNR configuration and non-chiral dispersion curves. For all panels dashed rectangles represent the unit cells.}
  \label{FIG:1}
\end{figure}

Nanoelectronics based on  graphene has  become a fast growing field \cite{neto2009electronic,geim:2009} with a number of technical applications \cite{wang:206803}. Electrons/holes low energy dynamics is described by the Dirac equation valid around the corners of the Brillouin zone (Dirac $K(K')$ points) \cite{brey2007eee,novoselov2005two,novoselov2007room}. One of the most striking consequences of the massless nature of the carriers is the so-called Klein paradox, which is the unimpeded penetration through high and wide potential barriers \cite{neto2009electronic,katsnelson2006chiral} as shown in Fig.\ref{FIG:1}. Graphene provides a convenient medium for experimental verification of the effect. The underlying physics of Klein tunneling is based on the notion of chirality as a symmetry between electrons and holes in graphene. Formally, each of the branches of the dispersion conical section is characterized by a specific sign of the pseudo-spin projection on to the momentum. The Klein effect is related to the conservation of this quantity across the potential barrier.
\par
In this paper, we explore the notion that chirality is necessary for Klein tunneling. For this one must compare transport in systems which provide either chiral or non-chiral massless Dirac fermions such as graphene nanoribbons (GNR). This are just an infinitely long carbon nanotubes unzipped along either armchair (ANR) or zigzag (ZNR) direction and then flattened. The ANR boundary conditions mix the valleys at $K\ (K^\prime)$ points giving the dispersion strongly depended on the number of carbon atoms ($N$) across the ribbon (modulo 3) \cite{zheng:165414} producing semi-metallic or insulating subbands. The fermions in the ANR are chiral and the potential barrier is perfectly transparent to electron scattering with angles close to normal incidence, in agreement with the Klein effect in the graphene sheets. This makes ANR a basis for graphene switches, random access memory and GNR field-effect transistors. When the electron motion is confined by zigzag edge, surface states describe the lowest energy bands of the ZNR providing the right (left) transmission channels \cite{wakabayashi2009edge,nakabayashi:066803} at $K\ (K^\prime)$ as shown in Fig.\ref{FIG:1}. Transmission through the barrier is possible if the gate potential acts as  an inter-valley scattering source. Unlike the ANR, the ZNR lowest energy dispersion does not show substantial dependence on $N$. However, the transmission drastically depends on this parameter. In the single channel regime, it has been predicted that the transmission is allowed (forbidden) for odd (even) $N$. This calls for referring to the case with odd number of atoms as anti-zigzag nanoribbon (AZNR) \cite{cresti:233402}. This current blocking was originally interpreted in analogy with the spin-valve effect in ferromagnetic junctions, thereby gaining the name of valley-valve effect \cite{rainis:115131,kinder-perfect}. Later, however, it was attributed to symmetric (asymmetric) coupling of states localized at opposite edges (sublattices) of the GNR \cite{li:206802}. Owing to the fact that the model for the ZNR and AZNR configurations are indistinguishable at the level of the Dirac equation \cite{akhmerov:205416}, the latter must be augmented with a pseudo-parity operator whose eigenvalues provide the correct dependence (in agreement with the tight-binding model) of the sublattice wavefunctions on $N$. Bellow, we show that when it comes to Klein tunneling in AZNR, pseudo-parity plays the role of the pseudo-spin in infinite graphene. Perfect transmission in case of head-on incidence is replaced by perfect transmission at the center of the ribbon. The chirality in AZNR is related to the projection of the pseudo-parity on the pseudo-spin around $K_x$ and $K^\prime_x$.
%----------------------------
%----------------------------
\par
The energy band structure of GNR depends on the shape of their edges. We shall focus on the ZNR configuration since the fermions in this geometry are not chiral \cite{akhmerov:205416}. At $K,\ (K^\prime)$ points, there is an excess right (left) moving conducting channel for $\pi (\pi^\star)$ bands.
In the spirit of single channel regime \cite{nakabayashi:066803}, these bands, and only them, will be considered further. Conventionally, these are given by $\mathbf{k}\cdot \mathbf{p}$ approximation for the tight-binding model \cite{neto2009electronic} yielding the following electron $+$ (hole $-$) dispersion valleys: $V(x)-E = \pm \sqrt{k^2_x-k^2_y}$.
Here the energy is measured in units of the hopping parameter $t = 2.7 \sqrt{3}/2  \; eV $ between nearest-neighbor carbon atoms and the wave vectors are in units of the inverse distance between them $a_0 = 1.42$ \AA.
The hard wall boundary conditions mix the transverse and the longitudinal electron momenta $k_x = k_y \coth {(W k_y)}$, where the GNR width is $W = 2 N -1 $ for even number $N$ of carbon atoms across the ribbon and $W = \tfrac{3}{2}(N-1)$, otherwise.
The chemical potential provided by the underlying metal contact (insulating strip, carbon nanotube) \cite{cheianov:041403,cayssol:075428} assumes the form of a sharp $n-p-n$ junction: $V(x) = V_0 \left[{\theta_+(x) - \theta_+(x-D)}\right]$. In this notation, $V_0$ is the height of the potential barrier and $\theta_+(x)$ is the Heaviside step function. In this way, we obtain the following equations:

As shown in Fig. \ref{FIG:1}, this alternating doping provides the scattering ($p-$ doped) region sandwiched in between left (right) leads ($n-$ doped).
The electric current through the potential barrier is given by the incident electron of energy $E$ in the range $\left[{0,\Delta}\right]$, where $2 \Delta$ is the energy separation between the top of the next highest valence and the bottom of the next lowest conduction band (parabolically shaped $\mathcal{V},(\mathcal{V}^\star)$ bands in Fig. \ref{FIG:1}).
To stay in the single channel regime, where the scattering occurs within lowest conduction $\pi$ and highest valence $\pi^\star$ bands, we require the barrier height $V_0$ in the range $\vert E-V_0 \vert \le \Delta$.
\par
The scattering problem with the sharp edge junction is fully determined by the wavefunction for the $\pi$ and $\pi^\star$ bands. The corresponding eigenfunctions for the two sublattices (A,B) can be factorized into the element-wise product of longitudinal, transverse, pseudo-parity and electron-hole parity two component wavefunctions as:
\begin {equation}
\label{eq:net_wavefunction}
 \Psi_{\pm K,N}(x,y) = \left({\begin{array}{c}
                    \Psi_A \\
                    \Psi_B
                  \end{array}}\right)= \phi_{\pm K}(x) \circ \chi(y)
 \circ \eta_{\pm K} \circ s
\end{equation}
Here, the longitudinal wavefunction depends on the parity $X_N$ of the  number of carbon atoms across the  ribbon:
\begin{equation}
\label{eq:longitudinal_wavefunction}
\phi_{\pm K,N}(x) = \frac{1}{\sqrt{2 \pi}}\left({\begin{array}{c}
                    \exp{\left({\pm i (K+k_x) x}\right)} \\
                    \exp{\left({\pm i (K+k_x) (x + X_N)}\right)}
                  \end{array}}\right)
\end{equation}
, where $X_{\rm even}=0$ and $X_{\rm odd} = \sqrt{3}/2$. Its presence stems from the suitable shape of the unit cells as depicted in Fig.\ref{FIG:1}. The choice of $\pm$ defines the right (transmitted) and the left (reflected) movers (direction of the electron group velocity for $E-V(x)>0$), respectively. The shape of the unit cells and, consequently, the form of the longitudinal wavefunction ascertains translational symmetry of the wavefunction \eqref{eq:net_wavefunction} along the ribbon. The transverse wavefunction component is
\begin{gather}
\label{eq:transverse_wavefunction}
\chi(y) =\left({\frac{2 k_y}{\sinh(2W k_y)-2W k_y}}\right)^{1/2} \left({\begin{array}{c}
                    \sinh \left[{k_y (W/2+y)}\right] \\
                    \sinh \left[{k_y (W/2-y)}\right]
                  \end{array}}\right)
\end{gather}
,which satisfies the Dirac equation at $\pm K = \pm 2 \pi / 3 \sqrt{3}$ as well as the hard-wall boundary conditions across the ribbon. In contrast to infinite graphene or the ANR, these bands describe localized bound edge states for $\Im m\ W k_y = 0$ and delocalized bulk states for $\Re e\ W k_y = 0,\; \Im m\ W k_y \le \pi$; while $\mathcal{V} (\mathcal{V}^\star)$ bands correspond to $\Re e\ W k_y = 0,\; \Im m\ W k_y > \pi$. It is interesting to notice that, in contrast to infinite graphene, the head-on incidence ($\Im m\ k_{y} \ll k_{x}$ occurring at $\Im m\ W k_{y} \to n \pi$) is prohibited due to mixing of the longitudinal and transverse wave vectors.
\par
There are two types of electron (hole) states in the ribbons given by the bonding and anti-bonding mixing of the sublattice wavefunctions. These strongly depend on the mirror symmetry (or lack of it) of the GNR, which follows from the tight-bing model \cite{cresti:233402}. To incorporate this feature into our model, we introduce the concept of pseudo-parity similar in nature to that in Ref. \cite{rainis:115131}:
\begin{equation}
\label{eq:pseudo_parity}
\eta_{\pm K} = \left({\begin{array}{c}
                    1 \\
                    (\pm K 3 \sqrt{3}/2 \pi)^N \exp{\left[{i \Im m\ \left({W k_y}\right)}\right]}
                  \end{array}}\right)
\end{equation}
The pseudo-parity component of the wavefunction \eqref{eq:net_wavefunction} determines wether the electron (hole) wavefunctions are either even or odd, forming bonding or anti-bonding states. Formally these can be defined by the following transformation \cite{wakabayashi2009edge}:
\begin{equation}
\label{eq:electron_hole_transformation}
\left({\begin{array}{c}
                    \Psi_b \\
                    \Psi_a
                  \end{array}}\right) = \frac{1}{\sqrt{2}}\left({\begin{array}{cc}
                    1 & 1\\
                    -i & i
                  \end{array}}\right) \left({\begin{array}{c}
                    \Psi_A \\
                    \Psi_B
                  \end{array}}\right)
\end{equation}
The pseudo-parity ensures alternation of the wavefunction sign for different bands as shown in Fig. \ref{FIG:1}. In fact, the parity of $N$ determines whether or not the bonding and anti-bonding wavefunctions switch between conduction and valence bands at $K\ (K^\prime)$ points. The electron-hole parity is comprised of the eigenvalues of the time-reversal operator:
\begin{equation}
\label{eq:electron_hole_symmetry}
s = \left({\begin{array}{c}
                    1 \\
                    \textrm{sgn}\left({E-V(x)}\right)
                  \end{array}}\right),
\end{equation}
where positive sign correspond to electrons in the lower conduction $\pi$ band, while negative sign stands for the holes in the highest valence $\pi^\star$ band.
\par
The problem of tunneling through the $n-p-n$ barrier can be described as follows. An incident particle belonging to the right moving channel ($K$ point) for $x<0$ strikes the potential barrier at $x=0$. The nature of the scattering determines the transmission probability density by employing continuity of the wavefunctions \eqref{eq:net_wavefunction} at the junctions and conservation of the pseudo-parity $\eta_{\pm}$. The reflected wave contributes to the left-moving channel (at $K^\prime$). It is clear that the necessary condition for non-zero transmission is the change of the wavefunction parity under reflection. This fails for $N$ even (ZNR configuration in Fig.\ref{FIG:1}) and the transmission vanishes identically $t \equiv 0$. The sublattice wavefunctions \eqref{eq:net_wavefunction} for $\pi$ band in the $n$-doped region and $\pi^\star$ band in the $p$-doped region have opposite sign for the pseudo-parity (opposite bonding) \cite{li:206802}.
The situation is drastically changed for AZNR ($N$-odd) configuration. The reflected wavefunction changes its pseudo-parity while the transmitted one keeps the parity of  the incident particle ($\pi$ band at $K$ and $\pi^\star$ at $K^\prime$ have the same pseudo-parity). The transmission probability density is given by
%\begin{widetext}
\begin{equation}
t(y) = \frac{\left({1-\beta}\right)^2 e^{i D (k_{x,2}-k_{x,1})} A}{\beta\left({e^{2 i D k_{x,2}}-1}\right) \left({\chi_{A,1}^2 \chi_{B,2}^2 + \chi_{A,2}^2 \chi_{B,1}^2}\right)-\left({2 \beta e^{2 i D  k_{
x,2}}+ \beta^2 +1}\right)A }
\label{eq:transmission_density}
\end{equation}

\begin{figure}[htbp]
  \centering
    \includegraphics[width=\textwidth]{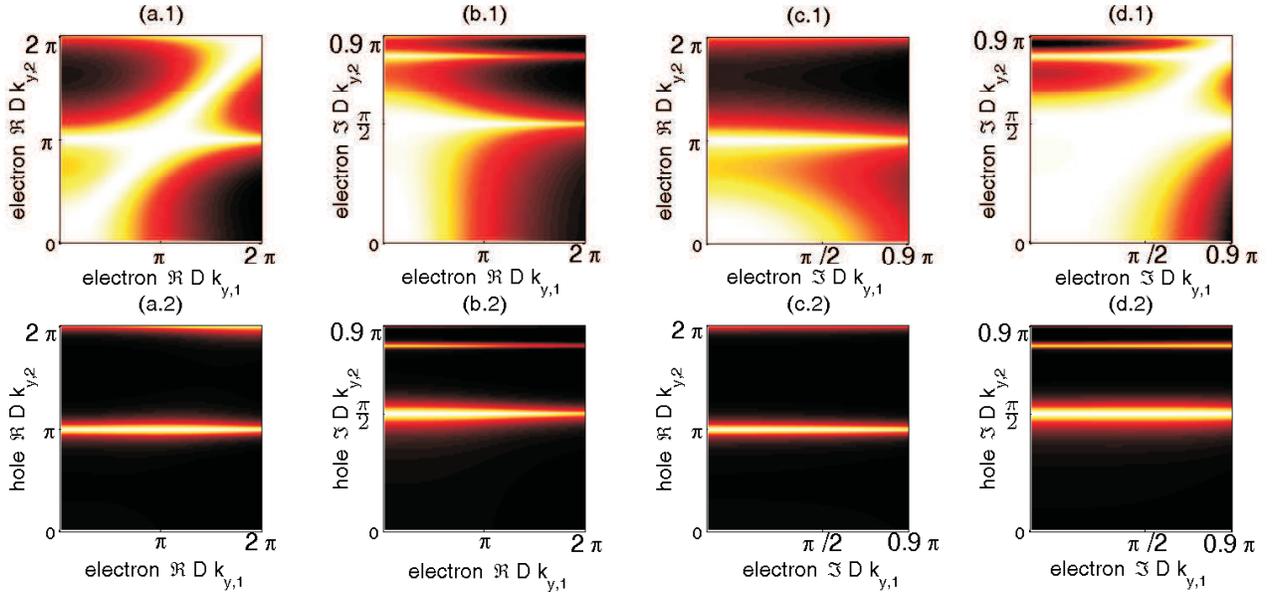}
  \caption{(Color online). The transmission $\vert{t(k_{y,1},k_{y,2})}\vert^2$ for 9-AZNR non-chiral $\pi$ electron. Panels (a.1)-(d.1) correspond to over the barrier transmission $E>V_0$. Panels (a.2)-(d.2) represent through the barrier $E<V_0$ transmission. $\Im m\ k_{y,1(2)}$ implies that the electron (hole) is in the bulk state, while $\Re e\ k_{y,1(2)}$ signifies the surface-bound states.}
  \label{FIG:2}
\end{figure}

\begin{figure}[htbp]
  \centering
    \includegraphics[width=0.5\textwidth]{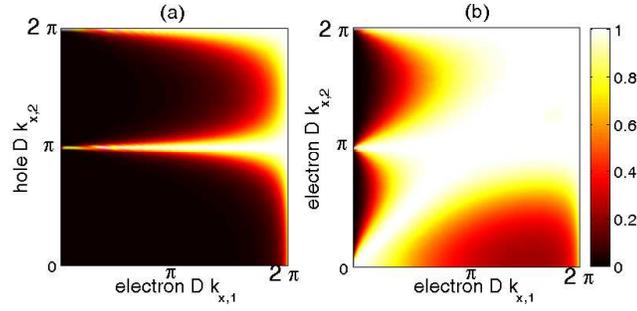}
  \caption{(Color online). The transmission $\vert{t(k_{x,1},k_{x,2})}\vert^2$ for an infinite graphene. Recall that conventionally the transmission is given in terms of the incident and refraction angles ($\phi_1,\phi_2$) between the transverse and longitudinal wave vectors in $n$- and $p$-doped regions correspondingly. The conversion is simply given by $\lambda k_{x,1} = 2 \pi \sin{(\phi_1)} \cot{\phi_2},\; \lambda k_{x,2}= 2\pi \cos {\phi_{1}}$. As in Ref.\cite{neto2009electronic}, the incoming electron wavelength is taken to be equal to the width of the $p$-doped region. Through the barrier $E \le V_0$ (a) and over the barrier $E \ge V_0$ (b) transmission of the chiral electron.}
  \label{FIG:3}
\end{figure}

Here, we introduced $\beta=\exp{\left({i \pi D X_N}/ \sqrt{3}\right)}$ and $A=s_{B,1} s_{B,2} \chi_{A,1} \chi_{A,2} \chi_{B,1} \chi_{B,2}$; the sub-indices $A,B$ on the wavefunction components denote the corresponding sublattices. The gated ($p$ doped) and $n$ doped regions are specified by the sub-indices $2$ and $1$ respectively, for instance $k_{x,2},k_{x,1}$ stand for the wave vectors along the ribbon in the $p$ and $n$ doped regions. The transmission probability (and hence, in the single-channel regime, the conductance \cite{akhmerov:205416} in units of $2 e^2/h$) is given by:
$\vert{t(k_{x,1},k_{x,2})}\vert^2=\frac{1}{W}\int \limits_{0}^{W} dy t(y)t^\star(y)$
and shown in Fig.\ref{FIG:2}. It jumps up to unity at $\Re e\ k_{y,2} D = n \pi$ for any integer $n$ in case of surface bounded holes, and at $\Im m\ k_{y,2} D = n \pi/2$ for the bulk-like holes. For the latter, there are additional resonances at $\Im m\ k_{y,2} D \to n \pi$. This situation corresponds to the quantum-dot-like states in the gated region \cite{trauzettel2007spin}. These resonances are similar to those in infinite graphene as shown in Fig.\ref{FIG:3}. The transmission is also ideal on the diagonal $k_{x,1} = k_{x,2}$ and corresponds to the absence of the gating potential.
\par
Above we analytically demonstrated that when the Dirac fermion picture is supplemented with the concept of the pseudo-parity for ZNR (AZNR) it removes the ambiguities of the ballistic transport and correctly describes the valley-valve effect (current blocking). We now turn to the role of pseudo-parity in the Klein effect in nanoribbons. As mentioned before, the Klein paradox in infinite graphene or metallic ANR is related to the conservation of pseudo-spin projection on the particle momentum, quantity known as chirality. That is the through the barrier unimpeded transmission occurs at the head-on incidence as indicated in Fig.\ref{FIG:3} (a) for $k_{x,1}D = 2 \pi$ for all values of the gate potential $V_0$ (any $k_{x,2}D$). Transformation from the sublattice basis to the bonding/anti-bonding representation yields that the head-on incident electron is in the purely bonding state ($\Psi_{a} = 0$) \cite{neto2009electronic}. However, the fermions in $\pi$ and $\pi^\star$ bands for ZNR and AZNR do not satisfy the conventional (pseudo-spin based) definition of chirality, thus often refereed to as non-chiral. Furthermore, the net transmission does not demonstrate such an effect. We now turn our attention to the transmission \emph{density} given by Eq. \eqref{eq:transmission_density}. It shows that at the center of the nanoribbon, transmission density is always perfect $t(y=W/2)=1$, regardless of the gate potential $V_0$. This effect is intimately related to the Klein paradox in  infinite graphene. Since, at the nanoribbon center, the incoming wavefunctions from the sublattices $A$ and $B$ form the bonding electron state ($\Psi_{a} (y=W/2) =0$). This holds true for both edge-bound and bulk states of the $\pi$ and $\pi^\star$ bands. It follows that we can relate the unimpeded transition density to conservation of the pseudo-parity rather than conservation of the pseudo-spin in infinite graphene. In AZNR, the pseudo-parity itself defines the chirality of the fermions. The valley-valve effect of the current blockage is the other manifestation of the Klein paradox through the pseudo-parity defined chirality in ZNR.
\par
It was also shown that with the use of density functional theory,  it is possible to obtaina more accurate band structure, which shows a gap in the ground state of zigzag nanoribon \cite{yao2009first}. It was also shown  by Son et al \cite{son2006half}  that the gap is inversely proportional
 to the width of the ribbon.   So, for a ribbon which is  wide  enough the bandgap becomes
negligibly small. In our model calculations, the pseudo-parity of the nanoribbon is more
 significant compared with intrinsic spin.  To keep the model simple and to focus on the leading effect
we have neglected spin altogether.
\par
In conclusion, the Klein paradox is unimpeded tunneling of the purely bonded Dirac electron state across arbitrary wide gated region in infinite graphene. Its another manifestation is perfect reflection in the graphene stacks. The paradox is conventionally interpreted in terms of the chirality conservation. The latter is defined as a projection of pseudo-spin on the direction of motion $s \cdot p$ for the two branches originated from the sublattices. In more general sense, the term chirality is often used to refer to an additional built-in symmetry between electron and hole parts of the spectrum. However, in chemistry, the term chiral is used to describe an object that is non-superposable on its mirror image. These two definitions cannot be related to infinite graphene since this system has no mirror reflection symmetry. The situation is not the same for the GNR. In accordance with the chemistry definition, ZNR's are non-chiral (the mirror plane is perpendicular to the GNR plane). On the other hand, AZNR is a chiral object in the chemistry sense. It is also chiral in the sense of pseudo-parity (two opposite projections of the pseudo-parity $\eta$ on the pseudo-spin $s$). As we have demonstrated, its conservation across the $n-p-n$ junction defines unimpeded electron tunneling at the AZNR center. Its conservation under the reflection is a source of valley-valve effect (perfect reflection) in ZNR. Both are facets of Klein paradox in nanoribbons.

\vskip 0.2 in
\noindent
{\bf Acknowledgments}:\
This work was supported by contract \# FA 9453-07-C-0207 of AFRL
and the Air Force Office of Scientific Research
(AFOSR).

\section{Appendix}
In this Appendix,  we consider the same problem for armchair nanoribbons. The wave function has the following form:
\begin{gather}
\Psi=\left({\begin {array}{c}
                    \Psi_A \\
                    \Psi_B
           \end {array}}\right)=\left({\begin{array}{c}
    \exp(i \textbf{K}\cdot\mathbf{r})\exp(i k_x x) \phi_A(\mathbf{r})+\exp(i \mathbf{K}'\cdot\mathbf{r}) \exp(i k_x x) \phi'_A(\mathbf{r}) \\
    \exp(i \textbf{K}\cdot\mathbf{r})\exp(i k_x x) \phi_B(\mathbf{r})+\exp(i \mathbf{K}'\cdot\mathbf{r}) \exp(i k_x x) \phi'_B(\mathbf{r})
    \end{array} }\right)
\end{gather}
\par
Here
\begin{gather}
\left( {\begin{array}{c}
           \phi_A \\
           \phi_B
           \end{array} } \right)= \circ \left({ \begin{array}{c}
                                                                   \exp(-i \phi) \exp(i k_n y) \\
                                                                   \exp(i k_n y)
                                                                   \end{array} } \right)
                                                                   \left( {\begin{array}{c}
           \phi'_A \\
           \phi'_B
           \end{array} } \right)= s \circ \left({ \begin{array}{c}
                                                                   - \exp(-i \phi) \exp(- i k_n y) \\
                                                                   \exp( - i k_n y)
                                                                   \end{array} } \right)
\end{gather}

\par
The wave function $\Psi$ must be continuous at $x=0$ and $x=D$ corresponding to  the boundaries of the region with potential $V(x)$.

\begin{gather}
\notag
\Psi(0,y;k_x,k_y;s_1;\phi)+r \Psi(0,y;-k_x,k_y;s_1;\pi-\phi)= a \Psi(0,y;q_x,k_y;s_2;\theta)+\\
\notag
+b \Psi(0,y;-q_x,k_y;s_1;\pi-\theta) \\
a \Psi(D,y;q_x,k_y;s_2;\theta)+b \Psi(D,y;-q_x,k_y;s_1;\pi-\theta)= t \Psi(D,y;k_x,k_y;s_1;\phi)
\end{gather}

In the  armcair case we have the same expressions connecting $k_x$, $k_n$, $\theta$ and $\phi$:
\begin{eqnarray}
\notag
\label{k_eq}
k_x=k_f cos(\phi)\; ;  \;
k_n=k_f sin(\phi) \\
q_x=\sqrt{(E-V_0)^2-k_{n}^2} \; ; \;
\frac{k_n}{q_x}=tan(\theta)
\end{eqnarray}
Also the phase for reflected waves is $\pi-\phi$ compared to the phase $\phi$ for incoming wave beacuse only this change of phase will ensure opposite $k_x$ with the same $k_n$ accourding to equataions \eqref{k_eq}
\par
In the simplest approximation when there is no $K$ and $K^\prime$ valley mixing, the above wave-function reduces to the four component vector

\begin{gather}
\Psi(\mathbf{r})=\left({\begin{array}{c}
   \exp(-i \phi) \exp(i k_n y) \\
   s \exp(-i k_n y)  \\
   - \exp(-i \phi) \exp(- i k_n y) \\
   s \exp(- i k_n y)
   \end{array}}\right) \ .
\end{gather}
\par
This gives rise to eight equations instead of the usual four which we need for obtaing the unknowns $r,a,b,t$. Our equations may be divided into four pairs, for which both equations in each pair are equivalent. Consequently, we have the same set of simultaneous equations to solve as for infinite graphene case, except $k_n$ accross the nanoribbon is quantized and given by

\begin{equation}
k_n= \frac{\pi\;n}{W}-\frac{2 \pi}{3 a_0} \ .
\end{equation}
A straightforward calculation leads to the coefficient for  transmission $t=t(\phi,\theta)$ is and hence the transmission probability given by \cite{neto2009electronic}

\begin{equation}
\label{Eq18}
T=t^\star t=\frac{cos^2(\phi)\;cos^2(\theta)}{\left[{cos(D q_x) cos(\phi) cos(\theta)}\right]^2+sin^2(D q_x) \left[{1-s_1 s_2 sin(\phi) sin(\theta)}\right]^2} \ .
\end{equation}
In the limit of a high potential barrier, i.e.,  $\vert V_0 \vert \gg  E$, the transmission \eqref{Eq18} will becomes

\begin{equation}
\label{19}
T=\frac{cos^2{\phi}}{1-cos^2(Dq_x)sin^2(\phi)}
\end{equation}
The case of special interest is $W=(3N+1)a_0$ which corresponds to the metallic subbands structure - there is no gap between the highest valence and the lowest conduction subbands.
In this case $k_n=0$ and as a result $\phi=0$ and the expression \eqref{19} will give $T \equiv 1$ - perfect transmission. For non-metallic case we have $sin(\phi)=\pi/k_f W$.

\bibliographystyle{unsrt}
\bibliography{bibliographene}

%--------------------------------
%-------------------------------

\end{document}